\documentstyle[epsf,aps]{revtex}
\begin{document}
\def\half{\frac{1}{2}}

\draft
\title{Band Structure and Transport Properties of CrO$_2$}

\author{Steven P. Lewis}
\address{Department of Chemistry, University of Pennsylvania,
Philadelphia, Pennsylvania 19104}
 
\author{ Philip B. Allen}
\address{Department of Physics, State University of New York,
Stony Brook, New York 11794-3800}

\author{Taizo Sasaki}
\address{National Researach Institute for Metals, 1-2-1
Sengen, Tsukuba 305, Japan}

\date{\today}

\maketitle

\begin{abstract}

Local Spin Density Approximation (LSDA) is used to calculate
the energy bands of both the ferromagnetic and paramagnetic phases
of metallic CrO$_2$.  The Fermi level lies in a peak in the
paramagnetic density of states, and the ferromagnetic phase
is more stable.  As first predicted by Schwarz, the magnetic
moment is 2 $\mu_{\rm B}$ per Cr atom, with the Fermi level
for minority spins lying in an insulating gap between oxygen $p$ and
chromium $d$ states (``half-metallic'' behavior.)
The A$_{1\rm g}$ Raman frequency is predicted to be 587 cm$^{-1}$.
Drude plasma frequencies are of order 2eV, as seen experimentally
by Chase.  The measured resistivity is used to find the
electron mean-free path $\ell$ which is only a few angstroms
at 600K, but nevertheless, resistivity continues to rise
as temperature increases.  This puts CrO$_2$ into the category
of ``bad metals'' in common with the high T$_{\rm c}$ superconductors,
the high T metallic phase of VO$_2$, and the ferromagnet
SrRuO$_3$.  In common with both SrRuO$_3$ and Sr$_2$RuO$_4$,
the measured specific heat $\gamma$ is higher than band theory
by a renormalization factor close to 4.

\end{abstract}


\section{Introduction}

The physical properties of rutile-structure oxides are
diverse, including insulators (TiO$_2$), antiferromagnets
(MnO$_2$), good metals (RuO$_2$), and metal-insulator systems
(VO$_2$).  Metallic CrO$_2$, with a Curie temperature $T_C \approx$
390K, is the only ferromagnet in this class.
Schwarz \cite{Schwarz} used LSDA band theory to predict
that the spin moment would be the full 2 $\mu_{\rm B}$
required by Hund's rules for the Cr$^{4+}$ ($3d^2$) ion. 
The Fermi level lies in a partly filled (metallic) band for the 
majority (up spin) electrons, but for minority (down) spins lies 
in a semiconducting gap which separates the filled oxygen $p$ 
levels from the Cr $d$ levels.  The situation where one spin species 
is metallic and the other semiconducting has been named 
``half-metallic'' by de Groot {\sl et al.} \cite{deGroot}.
Evidence of close to 100\% spin polarization was seen
in both spin-polarized photoemission \cite{Kamper}
and vacuum tunneling \cite{Wiesendanger}.  However,
this polarization was observed not at the Fermi energy,
but 2eV below.  Br\"andle {\sl et al.} \cite{Brandle93} found good agreement
between theory and experiment for the diagonal parts of the
optical conductivity tensor, but less good for the off-diagonal
magnetoconductivity properties.  These discrepancies have
not yet been resolved.

Besides the paper of Schwarz \cite{Schwarz}, LSDA band structure 
calculations have been reported by several other groups 
\cite{Mazin,Matar,Nikolaev}, with similar results.  We repeat 
these LSDA calculations using a plane-wave pseudopotential (PWPP)
method, and give both a prediction of the frequency of the $Q=0$ Raman 
active optic mode of A$_{1\rm g}$ symmetry and an analysis of transport
measurements using band-theoretical parameters.  This work is one of 
the first illustrations of the ability of a plane-wave pseudopotential 
technique to do a highly accurate LSDA calculation, and the first, to 
our knowledge, for a ferromagnetic compound.

\section{computational method }

The electronic ground state for CrO$_2$ is computed using the local spin density
approximation \cite{Gunnarson76} (LSDA) of density functional theory \cite{Hohenberg64}.
The form chosen for the exchange-correlation energy and potential is that of Ceperley
and Alder \cite{Ceperley80}.  We make the frozen-core approximation in which the core
electrons are fixed in their free-atom configuration, and only the valence electrons are
allowed to respond to the chemical environment of the solid.  Interactions between
the valence electrons and ion cores 
are described by efficient \cite{Troullier91} norm-conserving pseudopotentials 
\cite{Hamann79}.  The single-particle wavefunctions
are expanded in a plane-wave basis set which is well-converged for CrO$_2$ at a
cutoff energy of 81 Ry.

A worry with the PWPP method is that,
because spin-splitting of the exchange-correlation potential
is driven largely by the tightly-bound $d$-electrons, spin-polarizing effects would be 
prominent in the core region where the pseudo wavefunctions differ significantly from the
all-electron wavefunctions.   However, Sasaki, {\it et al.}, have recently demonstrated
\cite{Sasaki93} that this is not, in fact, a problem.  While the pseudopotential approach
does cause the 3$d$ wavefunction to shift in the core region relative to the all-electron
case, it also produces a similar shift in the spin-splitting of the exchange-correlation
potential.  Thus the total energy, which involves integration of the density
against the potential, is largely unaffected.
Sasaki, {\it et al.}, have also shown that it is important to employ the partial-core
correction scheme of Louie, {\it et al.} \cite{Louie82} in order to reduce core-valence
overlap errors introduced by nonlinearity of the exchange-correlation potential.
Spin-polarized calculations using this method have been performed for atomic and bulk
iron and nickel, and the agreement with all-electron LSDA calculations for structural,
electronic, and magnetic properties is outstanding \cite{Sasaki93}.

To our knowledge, the present study of CrO$_2$ is the first LSDA investigation of a
magnetic compound using the PWPP method.  
Results
for the magnetic and electronic structure obtained here are in excellent agreement
with previous all-electron LSDA band-structure calculations of this system 
\cite{Schwarz,Mazin,Matar,Nikolaev}, which
nicely illustrates the effectiveness and accuracy of this method.

\begin{figure}[t]
\epsfysize=3.75in\centerline{\epsffile{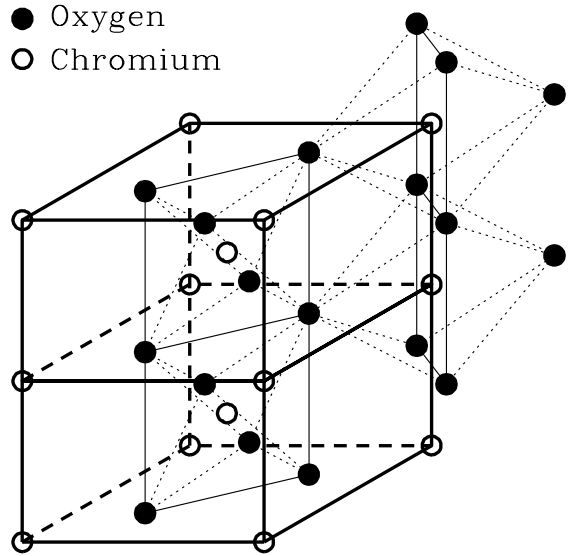}}
\caption{Illustration of the rutile structure.   Heavy solid and dashed lines 
demarcate unit cells of the crystal.   Thin lines emphasize the oxygen octahedra 
surrounding each chromium atom, with solid lines highlighting the equatorial planes 
and dotted lines linking them to the apical oxygen atoms.
\label{rutile}}
\end{figure}

The rutile structure (see Fig.\ \ref{rutile}) has a simple tetragonal Bravais
lattice with two formula units per unit cell.  It consists of chromium atoms octahedrally
coordinated by oxygen atoms, with the oxygen octahedra arranged in ``ribbons'' running
parallel to the tetragonal $c$-axis.  
Adjacent octahedra on the same ribbon share a common
edge, whereas octahedra on adjacent ribbons share a common corner and are situated
relative to each other according to a four-fold screw axis with non-primitive
translation equal to half the $c$-axis.  
Figure \ref{rutile} highlights the planar CrO$_4$
units which form the ribbons.  We refer to the ribbon oxygens 
as ``equatorial'' and the other two oxygens of the CrO$_6$ octahedra
as ``apical.'' The octahedra are orthorhombically distorted
away from the ideal geometry, with the apical oxygen atoms slightly more distant from
the central chromium atom than the equatorial oxygen atoms.  Chromium atoms are
located at the positions $[0,0,0]$ and $[\half,\half,\half]$ in lattice coordinates, and
the four oxygen atoms are located 
at $[u,u,0]$, $[1-u,1-u,0]$, $[\half+u,\half-u,\half]$,
and $[\half-u,\half+u,\half]$, where $u$ is a dimensionless internal coordinate.  The
measured values of the structural parameters, $a$, $c$, and $u$, are 4.419\AA, 2.912\AA,
and 0.303, respectively \cite{Porta72}.  In the present study, we have fixed the axes
at their experimental lengths and have opitimized the internal coordinate, $u$.  Its
computed value is 0.3043, in excellent agreement with experiment.  The irreducible
wedge of the tetragonal Brillouin zone has been sampled at 9 k-points for the purposes
of computing the self-consistent electron density.  The electronic density of states,
magnetic moment, Fermi energy, Fermi surface area, and transport coefficients are
determined using the linear tetrahedron method \cite{Lehmann72} on band energies
computed at 50 irreducible k-points.

\section{electronic structure}

\begin{figure}[t]
\epsfysize=4in\centerline{\epsffile{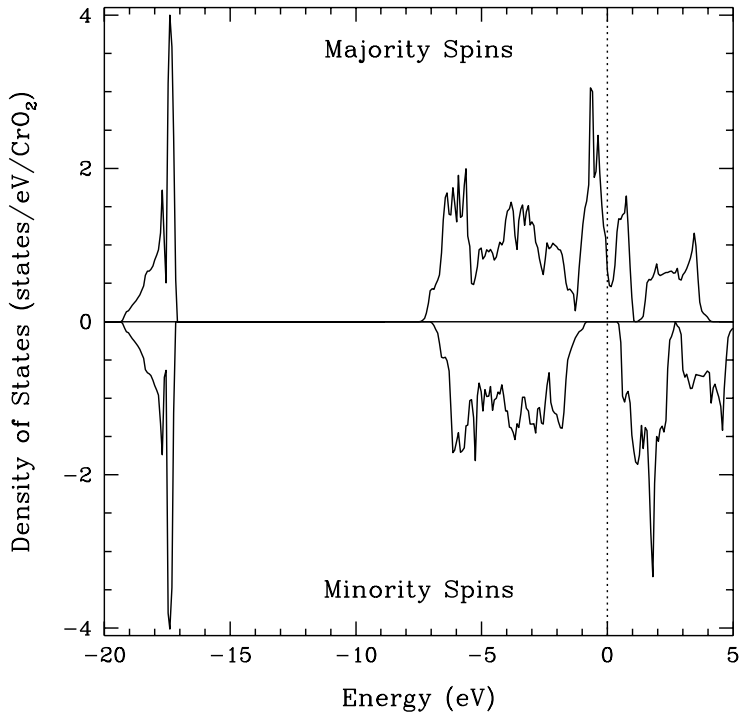}}
\vspace{6pt}
\caption{Densities of states for majority and minority spins in 
ferromagnetic CrO$_2$.   Majority (minority) states are plotted along 
the positive (negative) ordinate.  The dotted line denotes the Fermi 
energy.
\label{fm_dos}}
\end{figure}

Figure \ref{fm_dos} shows the computed electronic density of states (DOS) of
ferromagnetic CrO$_2$ for majority spins (positive axis) and minority spins (negative
axis).  These results 
agree well with earlier DOS calculations for this
material \cite{Schwarz,Mazin,Matar,Nikolaev}.  
Integrating the total DOS up to the total number of valence
electrons determines the Fermi energy, which is then used as the zero of energy.
The Fermi level intersects the majority spin bands near a local minimum in the DOS
and lies in a band gap of the minority-spin DOS.  Thus CrO$_2$ is half-metallic within
LSDA, as was first demonstrated by Schwarz \cite{Schwarz}.
This property leads to an integral
magnetic moment, which is found to be 2 $\mu_B$/CrO$_2$, the value predicted
by Hund's rules for the spin moment of the Cr$^{4+}$ ion.  
The density of majority states at the Fermi
level is 0.69 states/eV/CrO$_2$.

\begin{figure}[t]
\epsfysize=5in\centerline{\epsffile{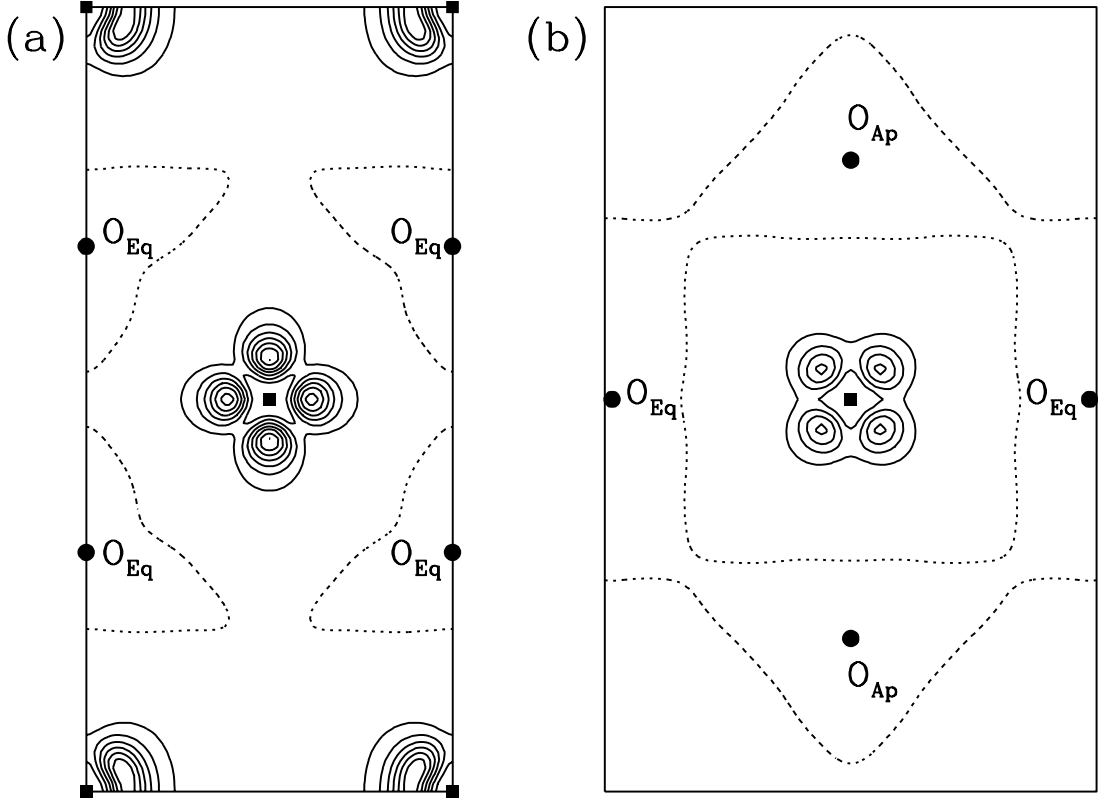}}
\caption{Spin density of CrO$_2$ in (a) the equatorial plane of the oxygen 
octahedron and (b) a plane of the octahedron containing two apical and two 
equatorial oxygens.  Adjacent contour levels are separated by 
30 $\mu_B$/unit~cell, with the zero contour designated by dashed lines.  
Chromium and oxygen atomic positions are designated by filled squares and 
circles, respectively.  Apical and equatorial oxygen atoms are distinguished 
by labels.
\label{spin_den}}
\end{figure}

Contour plots for the spin density 
($s(\vec{r}) = n_\uparrow(\vec{r}) - n_\downarrow(\vec{r})$) 
are shown in Fig.\ \ref{spin_den} 
for two cross-sections of the crystal.  Fig.\ \ref{spin_den}(a)
contains the $(1\bar{1} 0)$ plane formed by the equatorial 
oxygen atoms and the chromium atom.  
Fig.\ \ref{spin_den}(b)
shows a perpendicular plane containing two apical and two
equatorial oxygen atoms.  Both panels have
the same contour-level spacing.  These figures reveal that the 
spin density is highly localized around the chromium atom, 
and that it stems almost exclusively from the
chromium $3d$ states.  Fig.\ \ref{spin_den}(a) shows that the spin 
density has the ``cloverleaf'' shape of a $d_{xy}$ function, where a
local set of $x$, $y$, and $z$ axes are used with the origin on a Cr atom
and axes toward the octahedral O atoms, the axial oxygen defining
the local $z$-axis.  It is well-known that the crystal field
caused by an octahedron of surrounding negative charges 
splits the $d$ states of a transition metal ion,
with the three $t_{2g}$ states ($d_{xy}$, $d_{yz}$, and $d_{zx}$)
lying lower and the two $e_g$ states ($d_{x^2-y^2}$ and 
$d_{3z^2-r^2}$) lying higher.
Fig.\ \ref{spin_den}(b) shows that the $d_{xz}$,
or equivalently, the $d_{yz}$ component of $s(\vec{r})$ 
is also present with about half the strength of
the $d_{xy}$ component.  The $e_g$ states make no noticeable 
contribution.  

Bonding in CrO$_2$ has been
analyzed by Sorantin and Schwarz \cite{Sorantin} and by Burdette
{\sl et al.} \cite{Burdette}.  Referring to
the DOS in Fig.\ \ref{fm_dos},  the sharp band at about -17 eV 
arises from the oxygen $2s$ level, and is identical for 
both spin species, contributing nothing to the spin density.  
Similarly, the broad band from -7 to -1 eV comes
mostly from the oxygen $2p$ states, with a small admixture of 
chromium $3d$ character, and
is almost identical for majority and minority spins.
The source of the spin density is an exchange splitting of 
approximately 1.8 eV of the 
broad bands near and above the Fermi level.
These bands are primarily chromium $3d$ states of $t_{2g}$
parentage, with the $e_g$ bands higher in energy by the
crystal-field splitting of approximately 2.5 eV.
The exchange splitting shifts the minority spin $d$-bands 
above the Fermi level.   
The majority $t_{2g}$ band is two-thirds filled, with the
Fermi level lying in a ``pseudogap'' in the $t_{2g}$
manifold.  Following Sorantin and Schwarz, we
can explain the stability and the shape of the spin density in
CrO$_2$ as follows.  Only one of the three $t_{2g}$
states participates in covalent oxygen $p$-
chromium $d$ hybridization.  Hybridization creates both a bonding
and an antibonding hybrid orbital, with the bonding  orbital
reduced in energy, appearing in the occupied region of nominally
oxygen $p$ states, and the antibonding hybrid orbital remaining
in the chromium $t_{2g}$ manifold, but pushed up in energy
relative to the two non-bonding members, leaving the ``pseudogap''.
To see which $t_{2g}$ state hybridizes, consider an oxygen atom
with its three coplanar chromium nearest neighbors.  This
Cr$_3$O cluster builds $\sigma$-bonds from hybrids between the two
Cr $e_g$ states and two of the three O $p$ states.  (This hybridization
leaves antibonding $e_g$ states raised in energy relative to $t_{2g}$
states, and is part of the explanation of the crystal-field splitting.)
The leftover O $p$ state is the one which points perpendicular to
the Cr$_3$O plane.  This state hybridizes by forming a $\pi$-bond
with the Cr $t_{2g}$ state which is correctly oriented.
This state is orthogonal to the $d_{xy}$ state which lies in
the equatorial plane.  Thus one half of the $d_{yz}$ and $d_{zx}$
components of the $t_{2g}$ manifold is pushed upward by antibonding,
and this explains the dominance of $d_{xy}$ character in the 
spin density.

\begin{figure}[t]
\epsfysize=2.25in\centerline{\epsffile{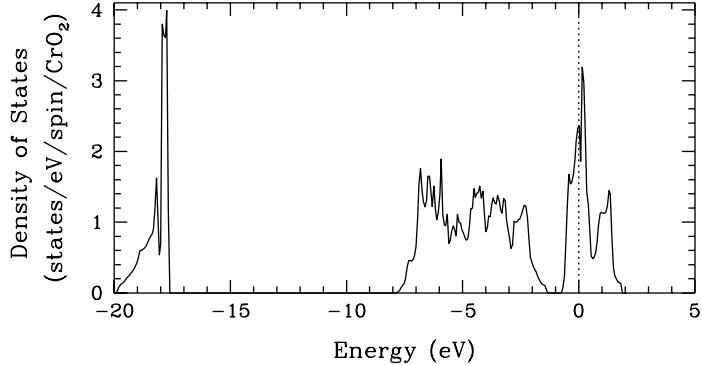}}
\caption{Density of states for paramagnetic CrO$_2$.  The dotted line 
denotes the Fermi energy.
\label{pm_dos}}
\end{figure}

For comparison purposes, we have also computed the band 
structure for CrO$_2$ without spin polarization.  The DOS 
for paramagnetic CrO$_2$ is plotted in Fig.\ \ref{pm_dos}.
It agrees fairly well with the LMTO calculation of Kulatov and Mazin \cite{Mazin},
although they did not find a pseudogap in the chromium $t_{2g}$ band.
Note that the Fermi level intersects a sharp peak of the DOS.  The
large DOS at the Fermi level (2.35 states/eV/CrO$_2$/spin) is an unstable,
high-energy situation, which is relieved by the formation of a ferromagnetic phase,
according to the usual Stoner argument.
The paramagnetic band structure also provides a useful comparison for understanding
the electronic transport properties of CrO$_2$, as is discussed later.

\section{Raman mode}

A symmetry analysis of the zone-center phonons of the
rutile structure (space group $P4_2/mnm$) shows that there
are four Raman frequencies and four infrared-active frequencies.
The Raman frequencies belong to irreducible
representations $A_{1g}$, $A_{2g}$, $A_{3g}$, and $E_g$,
whereas one of the IR
frequencies has $A_{4u}$ symmetry, with the rest having $E_u$ symmetry.

\begin{figure}[t]
\epsfysize=3.2in\centerline{\epsffile{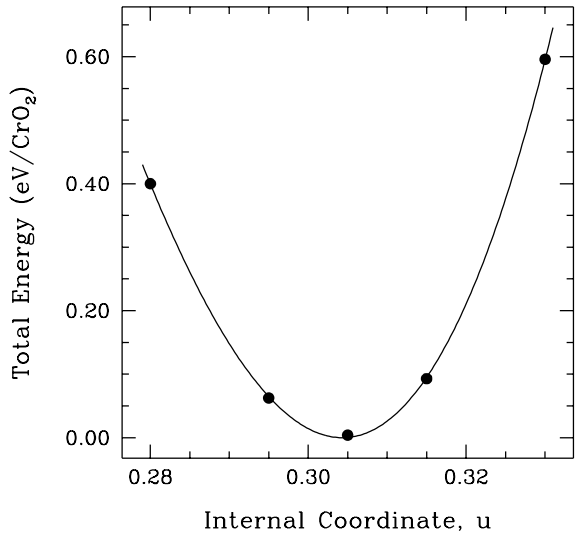}}
\caption{Total energy {\sl versus} internal coordinate $u$ for fixed
values of lattice parameters $a$ and $c$. 
\label{raman}}
\end{figure}

The $A_{1g}$ Raman mode is particularly simple, corresponding
to fluctuations of the internal coordinate, $u$.  
Thus, the energy of this mode is a by-product of
a calculation of the equilibrium structure.  
Figure \ref{raman} shows the variation of the total
energy of the crystal as a function of $u$, 
for fixed values of the lattice constants,
$a$ and $c$.  The solid line passing through the data 
points is a cubic-polynomial fit
to the data, with the zero of energy placed at the minimum.  
As mentioned above, the
equilibrium value of $u$ is predicted to be 0.3043.  
From the curvature of this function
at the minimum, we predict the frequency of the $A_{1g}$ 
Raman mode to be 587 cm$^{-1}$.
To our knowledge, this quantity has not yet been measured.

\section{Specific Heat}

The low temperature specific heat $C=\gamma T$ was 
measured \cite{Schubert} as $\gamma$=7 mJ K$^{-2}$ per mole CrO$_2$.
Using independent electron theory, $\gamma=(\pi^2 {\rm k_B}^2 /3)
N_{\gamma}(0)$, this corresponds to an effective density of
states $N_{\gamma}(0)$=3.0 states/eV per CrO$_2$ molecule.
This exceeds our calculated $N(0)$=0.69 by a factor 
$1+\lambda_{\gamma}$=4.3, a large enhancement.  Conventional
metals have an enhancement $1+\lambda$ from electron-phonon
interactions, and it would be possible to have $\lambda$
from this source fairly large, perhaps as big as 1, but not
as big as 3.3.  Analogous to the electron-phonon enhancement,
ferromagnetic metals can have a mass enhancement from
virtual exchange of spin fluctuations.  Because of the
absence of single particle states of opposite spin from
the spectrum at the Fermi energy, this source is expected
to be altered in a half-metallic material \cite{Irkhin}.
Perhaps one could patch up band theory in
some way, improving the LSDA exchange-correlation potentials 
or making self-interaction or gradient corrections, to diminish
the enhancement to a more reasonable value.  However, given
the reasonable success of LSDA in describing the total
energy and the qualitative nature of the spectrum of electronic
excitations farther away from the Fermi surface, it seems to
us more likely that the large specific heat enhancement
indicates a correlation effect beyond the reach of a
patched-up density-functional theory.  Very similar enhancements have
been reported in the ferromagnet SrRuO$_3$ \cite{sro}
and in Sr$_2$RuO$_4$ \cite{various}

\section{Resistivity}

\begin{figure}[t]
\epsfysize=4in\centerline{\epsffile{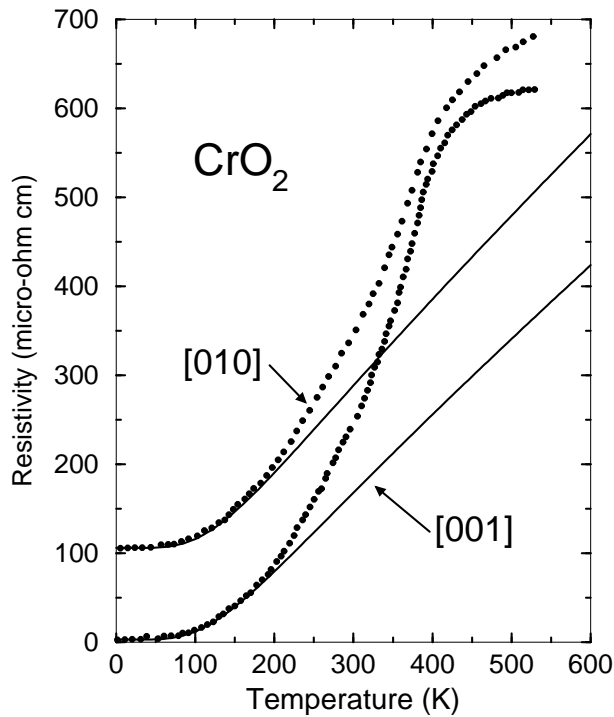}}
\vspace{12pt}
\caption{Resistivity of CrO$_2$ measured in ref. \protect\cite{Redbell}.
The data for the [010] direction have been shifted up by 100
$\mu\Omega$cm for clarity.  The solid curves are Bloch-Gr\"uneisen
functions fitted to the low T data as explained in the text.
\label{cro2res}}
\end{figure}

Transport properties of ferromagnets have been reviewed by
Campbell and Fert \cite{Campbell}.
The electrical resistivity of single crystal samples of CrO$_2$
was measured by Redbell {\sl et al.} \cite{Redbell}; the data
are transcribed on Fig.\ (\ref{cro2res}).
Similar results were reported by Chamberland \cite{Chamberland}.
The solid curves on Fig.\ (\ref{cro2res}) are Bloch-Gr\"uneisen fits
to the low-temperature data.

\begin{figure}[t]
\epsfysize=4in\centerline{\epsffile{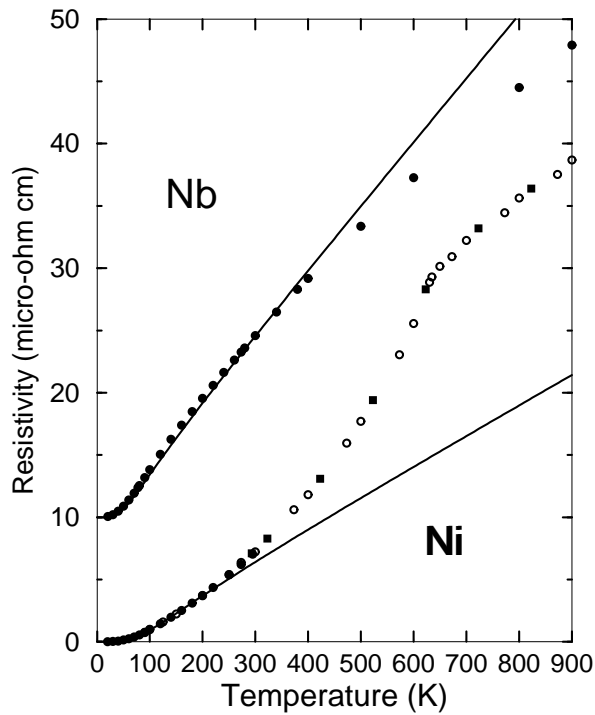}}
\vspace{12pt}
\caption{Resistivity of Ni and Nb {\sl versus} temperature
showing canonical behavior of ferromagnetic and nonmagnetic
metals.  The data for Ni are from Ref.\ \protect\cite{White}
(filled circles), Ref.\ \protect\cite{Laubitz} (open circles),
Ref. \protect\cite{Powell} (filled squares), and Ref.\
\protect\cite{Roeser} (open squares.)
The data for Nb, from Ref.\ \protect\cite{Abraham}, have been
shifted up by 10
$\mu\Omega$cm for clarity.  The solid curves are Bloch-Gr\"uneisen
functions as explained in the text.
\label{ninbres}}
\end{figure}

In order to interpret the resistivity curves, it is useful
to choose Ni and Nb as reference materials exhibiting canonical 
resistivity behavior.  The data, shown on Fig.\ (\ref{ninbres}), 
are from Refs.\ \cite{White,Laubitz,Powell,Roeser} 
for Ni and from Ref.\ \cite{Abraham} for Nb.
Boltzmann transport theory evaluates the current by
summing up the velocity of the occupied quasiparticle states.
Rather than solving the Boltzmann equation for the quasiparticle distribution
function, it is usually a reasonable approximation to use the {\sl ansatz}
that the distribution function $F$ is a displaced Fermi Dirac distribution,
$F_s(\vec{k})=f(\epsilon_s(\vec{k}+e\vec{E}\tau_s /\hbar))$.  
The spin state is labeled by $s=\uparrow {\rm or} \downarrow$.
This yields a formula for the conductivity,
\begin{equation}
\sigma_{xx}=e^2 N_{\uparrow}(0)\langle v^2 _{x\uparrow}\rangle\tau_{xx\uparrow}
   +e^2 N_{\downarrow}(0) \langle v^2 _{x\downarrow}\rangle \tau_{xx\downarrow}
\label{sigma}
\end{equation}
where the angular brackets signify a Fermi surface average
$\langle g \rangle \equiv \sum_k g(k) \delta(\epsilon_k)/\sum_k 
\delta(\epsilon_k)$.
The temperature dependence of the resistivity is controlled by the
scattering times $\tau_{\alpha\beta s}$.  For electron-phonon scattering,
this can be modelled by the Bloch-Gr\"uneisen formula,
\begin{equation}
\hbar/\tau_{xxs}=4\pi k_B T \int_0^{\omega_D} 
\frac{d\Omega}{\Omega} \alpha^2_{xxs}F(\Omega)
\left[\frac {\hbar\Omega/2k_B T}{\sinh(\hbar\Omega/2k_B T)}\right]^2
\label{tau}
\end{equation}
where the Debye model 
$\alpha^2_{xxs}F(\Omega)=2\lambda_{xxs}(\Omega/\omega_D)^4$
is used.  If the quasiparticle properties $N_s(0)$ and 
$\langle v^2_{xs} \rangle$ are known, then the parameters needed
to fit data are $\lambda_{xxs}$ and $\Theta_D=\hbar\omega_D /k_B$.

The simplest case to model is Nb, a cubic paramagnet, so that
all Cartesian directions and both spin directions have the 
same value of $\lambda_{xxs}$, denoted $\lambda_{\rm tr}$.  
It is convenient to define a ``Drude plasma frequency'' 
tensor $\Omega_P$ by
\begin{equation}
\Omega^2_{Pxx}=4\pi e^2\sum_s N_s(0)\langle v^2_{xs}\rangle
\label{plasma}
\end{equation}
Using $\hbar\Omega_P$=9.12eV for Nb as calculated by Papaconstantopoulos
\cite{Dmitri}, $\Theta_D=275$K from the low $T$ specific heat
\cite{Kittel}, and setting
$\lambda_{\rm tr}$ equal to the superconducting $\lambda$=1.05
found in tunneling \cite{Wolf}, the Bloch-Gr\"uneisen curve
shown in Fig.\ (\ref{ninbres}) follows with no adjustable
parameters.  A slightly better fit could be produced by reducing
the value of $\Theta_D$, which would accord with 
knowledge from tunneling about the actual effective phonon frequencies.
The fit is good up to 500K, beyond which the data
slowly turn downward below the theory, probably
signalling the onset of ``saturation'' \cite{Allen}.  
In many elemental d-band metals and intermetallic
compounds, when the electron mean-free path $\ell$ becomes less than 10\AA,
the thermal increase of resistivity is slowed.  When the mean free
path would hypothetically (according to Boltzmann theory) become
as short as 5\AA, the temperature dependence is almost completely
arrested.  Boltzmann theory is expected to fail when
$\ell$ is not large compared with a lattice spacing.  Typically
this occurs when the resistivity is within a factor of 2 of
100$\mu-\Omega$cm.  From the known parameters of Nb, it is predicted
that $\ell$ (defined as $\langle v^2\rangle^{1/2}\tau$)
at 500K should be 14.5\AA, close to (but a little higher
than) where ``saturation'' is expected.

An objection to the procedure used here is that Boltzmann theory
needs the velocities of the band quasiparticles, while the
single-particle energies $\epsilon_k$ calculated using LSDA band
theory are not guaranteed to have any physical meaning.
The only reliable test of this objection is comparison of
LSDA eigenenergies with experimental quasiparticle properties.
Tests for the resistivity of paramagnetic metals
have generally worked very well \cite{Allen1},
similar to the result seen here in Nb.

Nickel is a cubic ferromagnet ($T_C$=627K) so there are two electron-phonon
parameters, $\lambda_{\uparrow}$ and $\lambda_{\downarrow}$.
The resistivity cannot be fit with Bloch-Gr\"uneisen
curves; above 200K extra scattering occurs because of
fluctuating spins.  If we assume that below 200K electron-phonon
scattering dominates, we can still not determine separately all the 
constants of a Bloch-Gr\"uneisen curve, because at temperatures much
less than $\Theta_D$ only a single combination, $\lambda/\Omega_P^2\Theta_D^4$
enters the expression.  To get a rough idea, we make the model
$\lambda_{\uparrow}=\lambda_{\downarrow}$ and take $\Theta_D$=450K 
from low $T$ specific heat.  Using the joint up and down spin
Drude plasma frequency $\Omega_P$=6.96eV \cite{Dmitri}, the 
Bloch-Gr\"uneisen curve drawn on Fig.\ (\ref{ninbres}) corresponds 
to the choice $\lambda$=0.27.  This is a reasonable value, intermediate
between the values 0.13 and 0.47 found for the related metals Cu and Pd
\cite{Allen1}, and agrees with the result $\lambda$=0.24 found in
the pioneering calculation of Yamashita {\sl et al.} \cite{Yamashita}.  
It would also be reasonable to choose a lower value
of $\Theta_D$, obtaining a correspondingly reduced value of $\lambda$,
and a fit which follows the experimental $\rho(T)$ up to a somewhat
smaller temperature.  From the measured specific heat and the band values
of Papaconstantopoulos, one finds $\lambda_{\gamma} \approx 0.70$,
significantly bigger than can be reconciled with resistivity, and
suggesting that there is an additional source of renormalization 
beyond electron-phonon effects.

In principle, Boltzmann theory should still work for Ni at 
temperatures up to at least $T$=900K.
If only the electron-phonon mechanism were causing
scattering, then the value $\lambda$=0.27 implies that the mean-free
path for up spin electrons would be 35\AA and for down spin electrons
it would be 12\AA, using the Fermi velocities found by Papaconstantopoulos
\cite{Dmitri}.  However, the actual resistivities, because of spin
scattering, are higher by 1.72 than the electron-phonon fit, so the
corresponding mean-free paths are reduced to 20\AA and 7\AA.
But at 900K Ni is no longer a ferromagnet, and separate
up and down spin Fermi surfaces no longer exist.  Nevertheless,
the mean free paths suggested by this analysis are long enough that
it seems likely that the basic premises of Boltzmann theory are not
badly violated.  The fact that $\rho(T)$ continues to increase with
temperature at 900K suggests a metal where transport is still
governed by propagating quasiparticles which scatter more
rapidly at higher $T$.

We now apply this kind of analysis to CrO$_2$.
At low $T$, there are only majority (up) spins
at the Fermi surface, and thus only two electron-phonon
parameters, $\lambda_{xx}$ and $\lambda_{zz}$, which can be
separately fit to $\rho_{xx}$ and $\rho_{zz}$.  The Bloch-Gr\"uneisen
curves of Fig.\ (\ref{cro2res}) use $\Theta_D$=750K, taken
from low $T$ specific heat of the neighboring compound
TiO$_2$ \cite{Sandin}.  Using the calculated plasma frequencies
of Table I, we find $\lambda_{xx}\approx$ 0.8 and
$\lambda_{zz}\approx$ 0.9.  These values are a little higher
than found in typical oxide metals (compare $\lambda\approx$ 0.45
in RuO$_2$ \cite{Glassford}.)  The values of $\lambda$ are reduced
by a factor of 2 if we fit using $\Theta_D$=500K, which characterizes
the specific heat of TiO$_2$ in the range 15-65K \cite{Sandin}, and is an
equally reasonable choice.  The corresponding Bloch-Gr\"uneisen
curves deviate from the measured $\rho(T)$ at $T > 125$K
instead of $T > 190$K seen in Fig.\ (\ref{cro2res}).
In either case, because of the small Fermi velocities, the corresponding
$\ell$ values get small at higher $T$.

The mean free path has no unique definition.  Previously
we used $\ell=\langle \vec{v}^2\rangle^{1/2}\tau$, but an equally
good choice is $\ell=\langle \vec{v}^2\rangle\tau/\langle|\vec{v}|\rangle$.
In ferromagnetic CrO$_2$ we find 
$\langle \vec{v}^2 \rangle ^{1/2}=2.5\times 10^5$m/s and
$\langle |\vec{v}| \rangle =2.2\times 10^5$m/s, so the two
definitions differ by an unimportant amount.  Boltzmann theory
gives the following approximate relation:
\begin{equation}
\ell = \frac{\sum \vec{v}_k^2 \tau \delta(\epsilon_k)}
                 {\sum |\vec{v}_k| \delta(\epsilon_k)} \cong
\frac{(2\pi)^3 \hbar}{e^2} \frac{1}{A_{FS}^{\uparrow}+A_{FS}^{\downarrow}}
\left( \frac{1}{\rho_{xx}}+\frac{1}{\rho_{yy}}+\frac{1}{\rho_{zz}} \right)
\label{mfp}
\end{equation}
This provides a robust way to estimate mean-free paths, because
the only theoretical input required is the area $A_{FS}^s$ of the
Fermi surface.  Density functional theory has an excellent ability
to give the correct shape of Fermi surface, even under conditions
where the velocity of the quasiparticles at the Fermi surface, or
the density of states, may be off.  Our calculated areas are given
in Table I.   At low $T$, where $\rho$ is only a few $\mu-\Omega$cm,
$\ell$ in CrO$_2$ is hundreds of Angstroms.  At $T$=200K, we
estimate $\ell=36$\AA.  This is still the regime of a good 
band Fermi liquid.  At higher $T$, the resistivity increases rapidly,
and $\ell$ diminishes to $\approx 7$\AA at the Curie temperature,
390K.  This number is estimated assuming the ferromagnetic electronic
structure.  If instead we use the paramagnetic Fermi surface, the
area is much larger and values of $\ell$ are reduced by a factor
of 4.8 to sizes less than 2\AA.  This is the regime which has
been called a ``bad metal'' by Kivelson and Emery \cite{Kivelson}.  
In traditional
bad metals like A15 structure intermetallics (Nb$_3$Sn, etc.)
the resistivity is temperature-independent at this point, but in
``exotic bad metals'' like the high $T_c$ superconductors, 
resistivity continues to rise even though propagating quasiparticle
states of the Landau type cannot be present.  Earlier \cite{Schulz}
we found that the high $T$ metallic (rutile structure) phase of 
VO$_2$ was also a ``bad metal'' in the ``exotic'' sense, and also 
the ferromagnet SrRuO$_3$ \cite{sro}, so it should not be surprising 
if CrO$_2$ is also.

It is interesting to think about the nature of scattering in
CrO$_2$ at temperatures near $T_C$.  Figure (\ref{cro2res})
suggests that roughly half the scattering involves spin flips
caused by spin fluctuations.  The mean distance between
spin-flip scatterings is $\approx \surd 2\ell$, or 10\AA. 
A spin flip takes an electron
off the majority (up) spin Fermi surface into an insulating gap.
It cannot remain there long unless joined by a sufficient
number of other flipped electrons such that the local majority
has changed spin direction and the down spins are now metallic
with a Fermi surface leaving up spins insulating.  Thus it is
easy to imagine that the scattering is not completely incoherent,
but has a collective feature.  As temperature rises to near the
Curie temperature, the bulk magnetization decreases primarily through
fluctuating formation of opposite spin domains.  It seems natural to
associate the spin-flip scattering length with the size of these
domains.  Above $T_C$ the same picture can be used, except that
now the volume fraction of up and down domains is equal.  Within
each domain, electrons roughly behave as if they had the
ferromagnetic band structure, with majority spin electrons
having a Fermi surface, but badly lifetime-broadened.  With a 10\AA
spin-flip mean-free path, and a 7\AA total mean free path, the
local Fermi liquid picture is close to total destruction.  Looking
at the resistivity shown in Fig.\ (\ref{cro2res}), it is hard
to tell whether the resistivity is saturating or not.  
Additional measurements at higher $T$ would be interesting.

\acknowledgements

We thank R. J. Gambino and W. E. Pickett for useful advice.
This work was supported in part by NSF grant no. DMR-9417755, and
by the Division of Materials Sciences, U.\ S.\ Department
of Energy, under contract DE-AC02-76CH00016.  Computational
support was provided by the Pittsburgh Supercomputer Center.

\vspace{10pt}
\begin{table}
\caption{Measured, calculated, and derived properties of $\rm CrO_2$}
\vspace{9pt}
\begin{tabular}{ldd}
 Quantity & Ferromagnetic & Paramagnetic \\ \hline
  && \\
  \multicolumn{3}{c}{Experimental Values} \\
  && \\
  $a$ (\AA) (Ref.\ \cite{Porta72})                 & 4.419       &       \\
  $c$ (\AA)                                        & 2.912       &       \\
  $u$                                              & 0.303       &       \\
  $\gamma$ (mJ/mole K) (Ref.\ \cite{Brandle})      & 7           &       \\
  $\rho$(0K) ($\mu\Omega$cm)(Ref.\ \cite{Redbell}) & $\approx 5$ &       \\
  $\rho$(300K)                                     & 250         &       \\
  $\rho$(500K)                                     &             & 570   \\ 
  && \\ \hline
  && \\
  \multicolumn{3}{c}{Theoretical Values} \\
  && \\
  $u$                                              & 0.3043      &       \\
  $\omega$(A$_{1g}$) (cm$^{-1}$)                   & 587         &       \\
  N(0) (states/eV/spin/CrO$_2$)                    & 0.69        & 2.35  \\
  $\sqrt{\langle v^2_{Fx}\rangle}$  (10$^5$m/s)    & 1.38        & 0.62  \\
  $\sqrt{\langle v^2_{Fz}\rangle}$                 & 1.56        & 0.79  \\
  $\Omega_{pxx}$ (eV)                              & 1.91        & 2.22  \\
  $\Omega_{pzz}$                                   & 2.15        & 2.84  \\ 
  Fermi Surface Area (\AA$^{-2}$)                  & 8.86        & 21.43 \\ 
  && \\ \hline
  && \\
  \multicolumn{3}{c}{Derived Values} \\
  && \\
  $1+\lambda_{\gamma}=N_{\gamma}(0)/N(0)$          & 4.3         &       \\
  mean free path, $\ell$(5K) (\AA)                 & 700         &       \\
  $\ell$(300K)                                     & 14          &       \\
  $\ell$(500K)                                     &             & 1.3   \\
\end{tabular}
\end{table}


\begin{references}

\bibitem{Schwarz}    K. Schwarz,
                     J. Phys. F {\bf 16}, L211 (1986).

\bibitem{deGroot}    R. A. De Groot, F. M. Mueller, P. G. van Engen,
                     and K. H. J. Buschow,
                     Phys. Rev. Letters {\bf 50}, 2024 (1983).

\bibitem{Kamper}        K. P. K\"amper, W. Schmitt, G. G\"untherodt,
                        R. J. Gambino, and R. Ruf,
                        Phys. Rev. Letters {\bf 59}, 2788 (1987).

\bibitem{Wiesendanger}  R. Wiesendanger, H.-J. G\"untherodt, G. G\"untherodt,
                        R. J. Gambino, and R. Ruf,
                        Phys. Rev. Letters {\bf 65}, 247 (1990).

\bibitem{Brandle93}       H. Br\"andle, D. Weller, J. C. Scott,
                        J. Sticht, P. M. Oppeneer, and G. G\"untherodt,
                        Int. J. Mod. Phys. {\bf B7}, 345 (1993).

\bibitem{Mazin}         E. Kulatov and I. I. Mazin,
                        J. Phys. Condens. Matter {\bf 2}, 343 (1990).

\bibitem{Matar}         S. Matar, G. Demazeau, J. Sticht, V. Eyert, and
                        J. K\"ubler,
                        J. Phys. I France {\bf 2}, 315 (1992).

\bibitem{Nikolaev}      A. V. Nikolaev and B. V. Andreev,
                        Sov. Phys. -- Solid State {\bf 35}, 603 (1993).

\bibitem{Gunnarson76}   O. Gunnarson and B. I. Lundqvist, 
			Phys.\ Rev.\ B {\bf 13}, 4274 (1976).

\bibitem{Hohenberg64}   P. Hohenberg and W. Kohn, 
			Phys.\ Rev.\ {\bf 136}, B864 (1964);
			W. Kohn and L. J. Sham, 
			Phys.\ Rev.\ {\bf 140}, A1133 (1965).

\bibitem{Ceperley80}    D. M. Ceperley and B. J. Alder, 
			Phys.\ Rev.\ Lett.\ {\bf 45}, 566 (1980); 
			J. Perdew and A. Zunger, 
			Phys.\ Rev.\ B {\bf 23}, 5048 (1981).

\bibitem{Troullier91}   N. Troullier and J. L. Martins, 
			Phys.\ Rev.\ B {\bf 43}, 1993 (1991).

\bibitem{Hamann79}      D. R. Hamann, M. Schl\"uter, C. Chiang, 
                        Phys.\ Rev.\ Lett.\ {\bf 43}, 1494 (1979).

\bibitem{Sasaki93}      T. Sasaki, A. M. Rappe, and S. G. Louie, 
			Sci.\ Rep.\ RITU {\bf A39}, 37 (1993); 
			Phys.\ Rev.\ B {\bf 52}, 12760 (1995).

\bibitem{Louie82}       S. G. Louie, S. Froyen, and M. L. Cohen, 
			Phys.\ Rev.\ B {\bf 26}, 1738 (1982).

\bibitem{Porta72}       P. Porta, M. Marezio, J. P. Remeika, and P. D. Dernier, 
			Mater.\ Res.\ Bull.\ {\bf 7}, 157 (1972).

\bibitem{Lehmann72}     G. Lehmann and M. Taut, 
			Phys.\ Stat.\ Sol.\ (b) {\bf 54}, 469 (1972).




\bibitem{Brandle}       H. Br\"andle, D. Weller, S. S. P. Parkin,
                        J. C. Scott, P. Fumagalli, W. Reim, R. J. Gambino,
                        R. Ruf, and G. G\"untherodt,
                        Phys. Rev. B {\bf 46}, 13889 (1992).

\bibitem{Redbell}       D. S. Redbell, J. M. Lommel, and R. C. DeVries,
                        J. Phys. Soc. Japan {\bf 21}, 2430 (1966).


\bibitem{Schubert}	N. Schubert and E. Wassermann,
			(unpublished); data quoted in ref. \cite{Brandle}.

\bibitem{Sorantin}      P. I. Sorantin and K. Schwarz,
                        Inorg. Chem. {\bf 31}, 567 (1992).

\bibitem{Burdette}	J. K. Burdette, G. J. Miller, J. W. Richardson, Jr.,
			and J. V. Smith,
			J. Am. Chem. Soc. {\bf 110}, 8064 (1988).

\bibitem{Irkhin}        V. Yu. Irkhin and M. I. Katsnel'son,
                        Physics -- Uspekhi {\bf 37}, 659 (1994).

\bibitem{sro}		P. B. Allen, H. Berger, O. Chauvet, L. Forro,
			T. Jarlborg, A. Junod, B. Revaz, and G. Santi,
			Phys. Rev. B {\bf 53}, 4393 (1996).

\bibitem{various}	Y. Maeno, H. Hashimoto, K. Yoshida, S. Nishizaki, 
			T. Fujita, J. G. Bednorz, and F. Lichtenberg,
			Nature (London) {\bf 372}, 532 (1994);
			T. Oguchi,
			Phys. Rev. B {\bf 51}, 1385 (1995);
			D. J. Singh,
			{\sl ibid}, {\bf 52}, 1358 (1995).

\bibitem{Campbell}	I. A. Campbell and A. Fert,
			in {\sl Ferromagnetic Materials}, edited by
			E. P. Wohlfarth (North-Holland, Amsterdam, 1982),
			vol. 3, Ch. 9, pp747-804.

\bibitem{Chamberland}	B. L. Chamberland,
			Mat. Res. Bull. {\bf 2}, 827 (1967).

\bibitem{White}		G. K. White and S. B. Woods,
			Phil. Trans. Roy. Soc. {\bf A 251}, 273 (1959).

\bibitem{Laubitz}	M. J. Laubitz, T. Matsumura, and P. J. Kelly,
			Can. J. Phys. {\bf 54}, 92 (1976).

\bibitem{Powell}	R. W. Powell, R. P. Tye, and M. J. Hichmann,
			Int. J. Heat Mass Transf. {\bf 8}, 679 (1965).

\bibitem{Roeser}	W. F. Roeser and H. T. Wensel, 
			{\sl Temperature, its Measurement and Control
			in Science and Industry}, (Reinhold, New York,
			1941) p.1312.

\bibitem{Abraham}	J. M. Abraham, C. Tete, and B. Deviot,
			J. Less Common Metals {\bf 37}, 181 (1974).

\bibitem{Dmitri}	D. A. Papaconstantopoulos,
			{\sl Handbook of the Band Structures of Elemental
			Solids} (Plenum, New York, 1986).

\bibitem{Kittel}	C. Kittel,
			{\sl Introduction to Solid State Physics},
			7th Ed. (Wiley, New York, 1996).

\bibitem{Wolf}		E. L. Wolf, J. Zasadzinski, G. B. Arnold, D. F. Moore,
			J. M. Rowell, and M. R. Beasley,
			Phys. Rev. {\bf B 22}, 1214 (1980).

\bibitem{Allen}		P. B. Allen,
        		in {\sl Superconductivity in d- and f-Band Metals},
        		H. Suhl and M.B. Maple, eds.
        		(Academic Press, NY 1980) pp. 291-304.

\bibitem{Allen1}	P. B. Allen,
        		Phys. Rev. B {\bf 36}, 2920-23 (1987).

\bibitem{Yamashita}	J. Yamashita, S. Wakoh, and S. Asano,
			J. Phys. Soc. Jpn. {\bf 39}, 344 (1975).


\bibitem{Sandin}	T. R. Sandin and P. H. Keesom,
			Phys. Rev. {\bf 177}, 1370 (1969).

\bibitem{Glassford}	K. M. Glassford and J. R. Chelikowsky,
			Phys. Rev. {\bf B 49}, 7107 (1994).

\bibitem{Schulz}	W. Schulz, L. Forro, C. Kendziora, R. Wentzcovitch,
        		D. Mandrus, L. Mihaly, and P.B. Allen,
        		Phys. Rev. B, {\bf 46}, 14001-4 (1992).

\bibitem{Kivelson} 	V. J. Emery and S. A. Kivelson,
			Phys. Rev. Lett. {\bf 74}, 3253 (1995).

\end{references}
\end{document}